\documentclass[aps,pre,twocolumn,showpacs,superscriptaddress]{revtex4}

\usepackage{amssymb}
\usepackage{amsmath}
\usepackage{color}
\usepackage[british]{babel}
\usepackage[dvips]{graphicx}

\begin{document}

\title{Sensing viruses by mechanical tension of DNA in responsive hydrogels}

\author{Jaeoh Shin}
\affiliation{Institute for Physics \& Astronomy, University of Potsdam, D-14476
Potsdam-Golm, Germany}
\author{Andrey G. Cherstvy}
\affiliation{Institute for Physics \& Astronomy, University of Potsdam, D-14476
Potsdam-Golm, Germany}
\author{Ralf Metzler}
\affiliation{Institute for Physics \& Astronomy, University of Potsdam, D-14476
Potsdam-Golm, Germany}
\affiliation{Department of Physics, Tampere University of Technology, FI-33101
Tampere, Finland}
\thanks{E-mail: rmetzler@uni-potsdam.de}

\date{\today}

\begin{abstract}
The rapid worldwide spread of severe viral infections, often involving novel
modifications of viruses, poses major challenges to our health care systems.
This means that tools that can efficiently and specifically diagnose 
viruses are much needed. To be relevant for a broad application in local health
care centers, such tools should be relatively cheap and easy to use. Here we
discuss the biophysical potential for the macroscopic detection of viruses
based on the induction of a mechanical stress in a bundle of pre-stretched DNA
molecules upon binding of viruses to the DNA. We show that the affinity of the
DNA to the charged virus surface induces a local melting of the double-helix
into two single-stranded DNA. This process effects a mechanical stress along the
DNA chains leading to an overall contraction of the DNA. Our results suggest
that when such DNA bundles are incorporated in a supporting matrix such as a
responsive
hydrogel, the presence of viruses may indeed lead to a significant, macroscopic
mechanical deformation of the matrix. We discuss the biophysical basis for this
effect and characterize the physical properties of the associated DNA melting
transition. In particular, we reveal several scaling relations between the
relevant physical parameters of the system. We promote this DNA-based assay for
efficient and specific virus screening.
\end{abstract}


\maketitle

\section{Introduction}

Modern means of transportation, civil aviation traffic in particular, effect
extremely rapid global spreading of diseases \cite{dirk,davoudi}, contrasting
the much slower spreading dynamics by traveling fronts, for instance, during
the Black Death in Europe \cite{black}. Concurrently, new infectious diseases
keep emerging constantly, driven by human or ecologic reasons \cite{daszak},
while disease-causing microorganisms are developing various forms of multiple
drug resistance \cite{antibiotics}. This development puts considerable strain
on modern health care systems, requiring rapid, reliable, and specific diagnosis
of infectious agents.

There exist a number of modern techniques to detect viral and bacterial
pathogens. However, most of these methods require considerable time and
resources. Thus, for the detection of bacteria typical techniques include
the polymerase chain reaction and bacterial culture tests \cite{sibley}. For
viruses, on which we focus in this work, the detection methods include
electrochemical \cite{patolskyFCS}, optical \cite{brand-optical}, surface
plasmon resonance \cite{chinaSPR}, and biosensor devices \cite{turner96,schoen02}.
Some of them reveal a high single-virus sensitivity and high selectivity to the
virus type. 

Several successful examples of viral detection by bioanalytical chemistry setups
have indeed been reported. One of them is the macroscopic swelling of a polymeric
acrylamide hydrogel, that is cross-linked by specially designed and folded
ssDNA molecules, upon specific binding of influenza viruses \cite{wang13,liu12}.
As demonstrated in that experiment, upon binding of a specific H5N1 influenza
strain to the aptamer-containing hydrogel, which coats the surface of a quartz
crystal microbalance biosensor, the linkages of DNA aptamers incorporated in the
hydrogel become disrupted and the entire gel structure swells measurably. This
highly specific and sensitive detection method for aptamer-based sensors has a
number of advantages, as compared to the viral antibody-coated sensors, which are
widely implemented for rapid detection of viruses, see, for instance, the
assessment in Ref.~\cite{li11}. Biosensor setups implementing surface plasmon
resonance platforms based on antibody-antigen interactions were also shown to
be efficient for a selective detection of various influenza, hepatitis B, and
HIV viruses \cite{lee13}. Surface plasmon resonance detection of avian influenza
strains with a selectively-binding DNA aptamer immobilized directly on the sensor
surface was developed in Ref.~\cite{wang13-2}, combined with the a dot blot assay
for a visual detection of viruses in tracheal swab samples.

Applications for use in local health care centers or even as mobile diagnostic
tools require the miniaturization of the detection device, should provide a
real-time signal without a prior amplification step, be easy to use, guarantee
reproducible signals---and be relatively inexpensive. Physical instead of
biochemical and biological techniques may indeed lead the way towards novel
diagnosis methods. Thus, a relatively inexpensive and compact atomic force
microscope setup to quickly and relatively cheaply test for bacteria and their
response to drugs was recently proposed \cite{giovanni}.

\begin{figure*}
\begin{center}
\includegraphics[width=14.2cm]{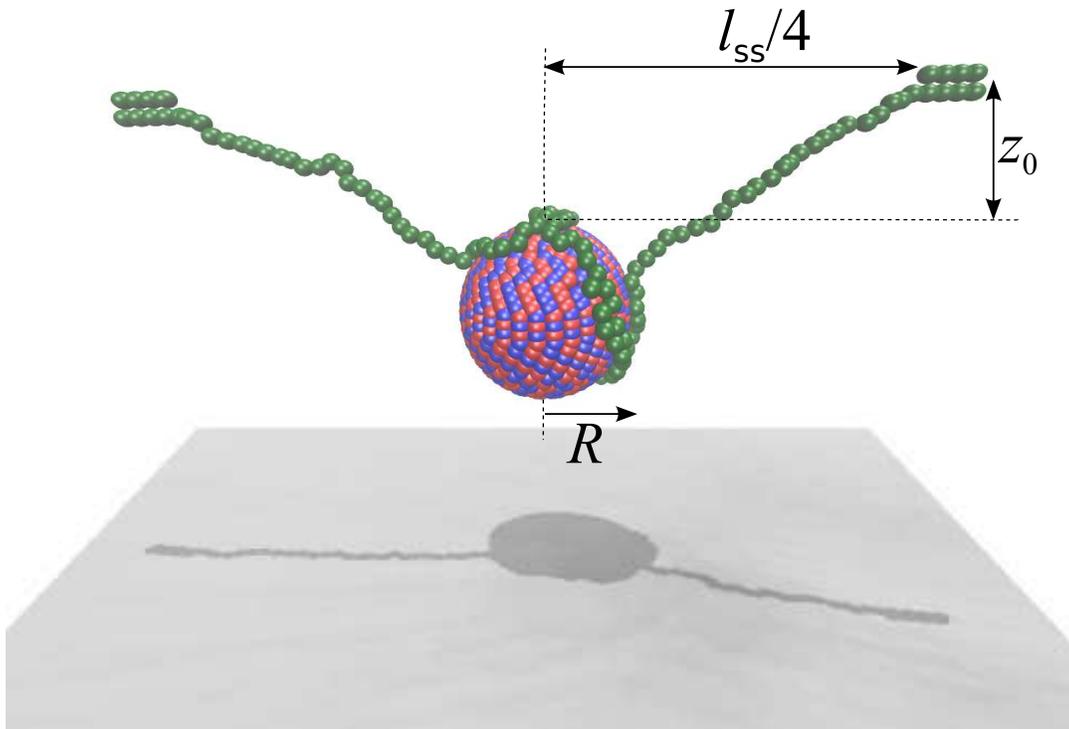}
\end{center}
\caption{Schematic of ssDNA-virus binding with the definition of model
parameters. Red and blue monomers on the shell correspond to attractive and
neutral patches on the "raspberry" surface of the viral capsid. We show one
of two ssDNA fragments of a partially denatured, linear DNA (green), at its
extremities we indicate intact dsDNA. A real chain
configuration as obtained in simulations for $l_{ss}=100 \sigma$, $p=1/2$,
$R=6\sigma$, and $\epsilon_A=4k_BT$  is shown. This corresponds to a strong
adsorption limit with a progressive wrapping of the flexible ssDNA chain around
the attractive sphere. The shell is composed of a spiral running from the
pole of the sphere along its surface, with red-blue monomers  positioned
according to the fraction $p$. See text.
\label{fig1}}
\end{figure*}

Could we come up with a method to detect viruses via a physical signal, for
instance, a mechanical contraction of a matrix such as a responsive
polyelectrolyte hydrogel, simply based on molecular interaction of the matrix
with the surface of the virus shell? In what
follows, from extensive computer simulations and statistical mechanical
calculations we demonstrate that viruses may cause a partial melting of
double-stranded DNA (dsDNA) molecules due to the binding affinity of the viral
shell for single-stranded DNA (ssDNA), see the illustration in Fig.~\ref{fig1}.
This partial melting of the double-helix leads to an overall contractile,
entropic force between the two extremities of the linear DNA chain of the order
of several pN. Our results demonstrate that the reaction of viruses with DNA
chains embedded in a hydrogel indeed leads to a significant, detectable
contraction of the hydrogel. We therefore promote such DNA based technology for
the rapid and inexpensive detection of viruses.

In our analysis we show that there exists a sharp phase transition of the
underlying DNA melting, as function of the binding affinity of ssDNA to the
surface of
the virus. We establish a phase diagram for the denaturation and reveal a scaling
relation for the fraction of intact base-pairs as function of the
ssDNA-virus binding affinity. Finally we obtain the contractile force per
virus-DNA pair as function of the binding affinity. Upscaling of these results
leads us to our central finding of the effective contraction of a hydrogel
equipped with a bundle of DNA chains arranged in parallel in the presence of
viruses.

\section{Virus-induced DNA melting and tension formation}

In the established thermodynamic models of DNA melting an alternating sequence
of double-helical and denatured segments of different length form a partially
molten DNA \cite{ps66,kamen,santa,kafri,everaers}. Once
a denaturation bubble is nucleated after overcoming a free energy barrier $F_s$
associated with the cooperativity parameter $\Omega=e^{-F_s/(k_BT)}\approx10^{
-5}$, base-pairs unzip (and zip close again) sequentially. DsDNA can be
stabilized by some DNA binding cations (protamine, alkaline earth cations) and
large ligands \cite{mcgee}, cationic lipids and surfactants, crowding agents
such as Ficoll-70 \cite{ficollTM}, basic poly-peptides, and DNA-binding proteins
(histones), as well as DNA-DNA attractive interactions \cite{acak05}. Conversely,
dsDNA is destabilized by DNA unwinding proteins (gene 32 protein) \cite{p53},
external DNA twist and super-coiling \cite{metzsuper,mukamel}, and stretching
\cite{rief,metz08}, as well as by single-strand DNA binding proteins and
intercalators \cite{williams}.

The generation of an entropic force upon melting of pre-stretched, linear DNA
fibers was demonstrated previously. Thus, for highly oriented dense DNA fibers,
pre-stretched by picoNewton forces, the thermally or chemically induced melting
transition of double-helical DNA was shown to trigger macroscopic changes of the
fiber length. This melting-induced fiber contraction can reach up to 70-90\% in
such mechano-chemical studies, solely due to the entropically favored shrinking
of the molten segments \cite{nordsmono1,nordsmono2,nords94}: the highly flexible
ssDNA segments have a much smaller equilibrium end-to-end distance compared to
the much stiffer dsDNA fragments \cite{grosberg}. Depending on the exact
conditions of the solution (salt concentration, temperature, etc.) and the
degree of torsional freedom, DNA entropic forces can indeed reach 10 to 40 pN
per single DNA molecule in the fiber.

How do viruses come into play? The electrical charge of viruses was already
studied in the 1920ies \cite{bedson}. Today we know that the external surface
of the viral capsid, the protein-based shell of viruses, for a number of virus
species features strongly non-uniform distributions of electric charges
\cite{rudi,viperDB,rudi13virus,cher11}. DNA binding to such shells can thus be
based purely on electrostatic interactions. For instance, upon infection by an
influenza virus, highly cationic haemo-glutinin HA glycoproteins domains, that
form protrusions on the viral surface, anchor to negatively charged sialic acid
receptors on the host cell surface \cite{flu1,flu2}. A number of modern anti-flu
drugs impede the viral infection proliferation via preventing these electrostatic
contacts from forming.

In such a scenario, the DNA-virus binding affinity, the elastic parameters of
DNA, and the energy difference between the molten and double-helical states of
DNA are all delicately sensitive to the salt concentration and temperature of
the ambient solution. For instance, in the range of $0.01$ to $0.2$ Molar of
monovalent salt the DNA melting temperature $T_m$ exhibits a logarithmic
dependence on the salt molarity of the solution \cite{meltsalt}, leveling off
at very large salinities \cite{meltsalt2}. The electrostatic repulsion of
interwound ssDNA strands becomes more pronounced at lower salt, thus
effectively reducing $T_m$. The effects of electrostatic interactions onto the
thermodynamics of the DNA melting transition near a charged interface were
analyzed in Refs.~\cite{petit11,fuchs}.

Without external forces, the electrostatically driven adsorption-desorption
transition of unconstrained polyelectrolyte chains such as DNA onto oppositely
charged curved surfaces has been examined in detail theoretically
\cite{acrw11,acrw13} and by computer simulations \cite{bach13}. The scaling
laws for the critical adsorption transition were obtained experimentally as
functions of polyelectrolyte chain-surface adhesion strength, salt conditions,
surface curvature, and chain stiffness \cite{dubin13}. To tackle such a
transition for patchy curved surfaces and adsorbing polymers, as considered here
numerically, remains a challenge.

Apart from these solution-sensitive, direct electrostatic attractions between
DNA and the surface of the virus, there exists an alternative method to effect
binding between ssDNA and the virus capsid. Namely, functionalized, chemically
engineered viral shells can bind both dsDNA and ssDNA with different propensity
via chemical linkers. For instance, ssDNA can bind better to viral shells covered
with ssDNA-binding proteins, as used by viruses to dock to their host cells
\cite{flu1,flu2}. Thus, by various chemical and biochemical methods, the binding
of ssDNA to virus shells may be effected experimentally. Using biochemical
linkers, the ssDNA binding may indeed be rendered specific to certain virus
types such as influenza, norovirus, HIV, or herpes. At the same time the binding
affinity
may reach relatively high values such that the bonds will not dissociate over
fairly long timescales.

The presence of viruses can thus favor the (partial) melting of DNA chain. How
can we combine the DNA-virus system with a hydrogel matrix? Hydrogels
\cite{okay00,okay09,russians} are extensively used in both industrial and
medical applications, for instance, for tissue engineering purposes. These
cross-linked polymeric materials feature a highly responsive behavior to
various external stimuli such as temperature \cite{okay13}, solvent quality,
pH as well as for detection of various substances
\cite{norway,mexico,wisch10,suhor10}. The volumetric changes of up to 100 times
in some hydrogels are reported \cite{okay09} due to electro-osmotic swelling
in low-salt solutions (cation accumulation). Viscoelastic hydrogels such as
agarose feature a number of rubbery characteristics \cite{hdelast1}.

\begin{figure}
\begin{center}
\includegraphics[width=8cm]{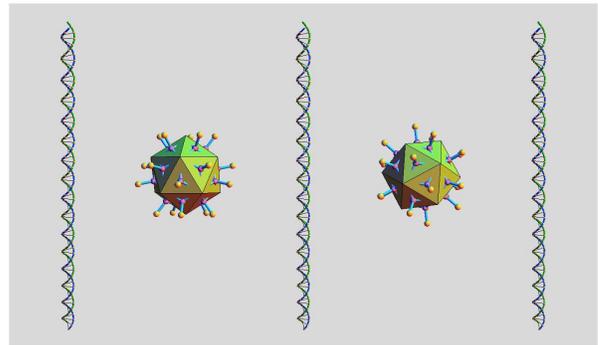}
\end{center}
\caption{Schematic of the parallel arrangement of DNA chains with centered
virus particles. Thermodynamically, the presence of the virus particles
triggers the partial melting of the DNA chains, leading to configurations as
shown in Fig.~\ref{fig1}.
\label{fig1a}}
\end{figure}

We assume that in the hydrogel individual DNA chains are supported in a
pre-stressed, almost linear configuration, and that they are aligned in parallel,
as sketched in Fig.~\ref{fig1a}. Perpendicular to their longitudinal axis, the
DNA chains occupy a square or hexagonal lattice. We further assume that after
adding viruses to the solution,
each square or triangular ``elementary cell'' contains a virus particle. Thus,
on average each virus is associated with one to two DNA double-helices, or two
to three ssDNA strands after partial denaturation of the DNA chain, compare
Fig.~\ref{fig1}. Given the
realistic parameters we use in our quantitative analysis below, this means that
each ssDNA strand has sufficient accessible area to bind to the surface of the
virus capsids, without significant overlap with competing ssDNA segments. In
what follows we consider the interaction of a single dsDNA strand with a given
virus capsid. Due to the approximate independence of ssDNA binding to the virus
capsids in all cells, we will then be able to upscale our results for the entire
DNA bundle-virus assembly in the hydrogel. In this approach the position of the
virus is fixed, reflecting the symmetry of the DNA-capsid configuration. We find
that viral particles in the hydrogel-DNA system destabilize dsDNA and effect a
macroscopic contraction of the hydrogel matrix.

\section{Results}

\subsection{DNA melting effects contraction of the hydrogel}

As shown in our model configuration in Fig.~\ref{fig1} the surface of the virus
is the distance $z_0$ away from the axis connecting the two end-points of the DNA
that are fixed in the hydrogel. The separation between the axis of
the DNA molecule from the center of the virus is thus given by $\frac{s}{2}=(z_0+
R)$, where, for simplicity, the virus capsid is taken to be spherical with radius
$R$, and that it is centered with respect to the longitudinal mid-point of the DNA
chain between its two extremities incorporated in the hydrogel (Fig.~\ref{fig1a}).
In our DNA-virus
pair, due to the binding affinity of ssDNA to the capsid the part of the
double-helix close to the virus capsid will melt (Fig.~\ref{fig1}). The entire
DNA molecule is thus composed of two dsDNA segments and two ssDNA segments
resulting from the molten middle part.

For now we concentrate on one of these ssDNA segments, whose
length $l_{\text{ss}}=l_b+2l_f$, in turn, is made up of the portion $l_b$ bound
to the viral shell and the two vicinal, unbound ssDNA fragments of length $l_f$
each of which connects the bound portion with the two intact dsDNAs. Once such a
configuration is established, the resulting entropic strain will be transferred
to the end-points of the DNA chain. Since the base-pair to base-pair distance is
approximately 0.34 nm in dsDNA, and the phosphate-phosphate distance 0.7 nm in
ssDNA, the maximal ssDNA length available for binding to the virus is
approximately two times the length of the corresponding dsDNA, provided that the
dsDNA fragments can untwist freely upon melting which may be achieved by the
specific anchoring in the hydrogel or by adding a nick close to the extremities
of the DNA chain.

In our simulations, already denatured ssDNA chains are equilibrated and their
most probable configurations are analyzed. We distribute positive electric
charges on the capsid such that they occupy an area fraction $p$. The
interaction free energy
$F$ with the virus capsid due to electrostatic binding between the DNA chain
and the positively charged sites on the capsid as well as the resulting
contraction force $f$ acting on the chain ends is computed, for different
virus radii $R$, DNA separations $z_0$ from the virus surface, and ssDNA-virus
attraction strengths $\epsilon_A$. This attraction strength is measured for one
monomer of the ssDNA chain that corresponds to a sphere of diameter $\sigma=4
\text{nm}$ accommodating $\approx12$ base-pairs of the dsDNA or 6 unpaired
bases of the ssDNA. The DNA melting transition necessary for ssDNA formation,
triggered by the free energy gain from binding to the viral surface, is analyzed
below as a function of these parameters.

\subsection{Free energy, forces, and adsorption transition}

The relaxation time to reach the equilibrium increases with the length of the
simulated ssDNA and for decreasing ssDNA-virus attraction strength. For an
ssDNA chain consisting of $n=101$ monomers the equilibration takes typically
$\approx10^6$ simulation steps, corresponding to approximately 10 min on a 3 GHz
workstation. For polymer chains of 201 and 401 monomers, the equilibration
time takes $\approx3\times10^6$ and $\approx10^7$ simulation steps. Every point
in the figures below is calculated as an average over at least $10^4$ polymer
configurations, after equilibration of the system.

It is natural to expect that for large DNA-virus capsid binding strengths the
thermal fluctuations of both the ssDNA and dsDNA chain domains become suppressed.
They are effectively pulled out by the binding-mediated force, $f_A$, directed
towards the viral particle. For different virus dimensions and DNA-virus
separations, this force contains a geometric correction factor $C$ and turns out
to follow the simple law
\begin{equation}
\label{force-contraction}
f_A=C\epsilon_A/\sigma,
\end{equation}
see below for details. Appendix \ref{appa} contains further information on
the model potentials used in the simulations that act between the chain monomers
as well as between the polymer chain and the virus surface.

The total free energy of a partially molten DNA double helix is comprised of the
sum of entropic and energetic contributions along the equilibrated chain, the
interaction energy of the polymer chain with the virus surface $U$, and the DNA
melting free energy $F_{\text{melt}}$. The DNA-virus attraction energy scales
with the length of bound ssDNA
\begin{equation}
U=B(p)\epsilon_Al_{b}/\sigma.
\end{equation}
The constant $B(p)$ accounts for multiple contacts that the chain monomers can
establish with the attractive sphere monomers, where $p$ quantifies the surface
fraction of attractive binding spots on the virus, see below.
The DNA melting free energy consists of the free energy cost for base-pair
unstacking, $\Delta F$, and the initiation free energy $F_s$ for nucleation
of a denaturation bubble,
\begin{equation}
F_{\text{melt}}=F_s+\Delta Fl_{\text{\text{ss}}}/(6\sigma).
\label{fmelt}
\end{equation}
For simplicity, all the DNA-related energies are assumed to be independent of
the DNA base-pair sequence, its GC content \cite{indianmelting}, and, most
crucially, the length of the ssDNA fragments. In the simulations presented below
we use $\Delta F=0.3-0.5k_BT$ and $F_s=10k_BT$. The relatively small $\Delta F$
values mimic a system away from equilibrium, where the values of $\Delta F\sim
1k_BT$ are typically used.

The denaturation equilibrium of dsDNA and the fluctuation spectrum of ssDNA
become altered by the ssDNA binding to the viral capsid. The optimal length of
the molten DNA $l_{\text{ss}}^{\text{opt}}$ is determined self-consistently
from the minimum of the total free energy functional. As order parameter for the
DNA adsorption transition we use the fraction of DNA base-pairs adsorbed to the
viral shell,
\begin{equation}
\theta=l_b/l_{\text{ss}}.
\end{equation}
The position of the phase transition boundary crucially depends on the
strength of the DNA-virus attraction,
the DNA density in the bundle, scaling as $1/[\pi(s/2)^2]$, and the virus
dimensions, $R$. It will be instructive to account for the maximal fraction of
ssDNA, $\theta_{\text{max}}$, to be adsorbed due to geometrical constraints of
the system, see below. In this state, the flanking ssDNA fragments are fully
stretched (strong adsorption limit). The value of $\theta_{\text{max}}$ is a
function of the virus size and the DNA-capsid separation.

We implement the (standard) dynamical criterion for adsorption of a ssDNA chain
monomer to the virus. When a monomer stays in contact with the surface for more
than 50\% of the simulation time, it is considered adsorbed. We assume that a
portion $0<p<1$ of the spherical virus surface, composed of monomeric patches as
indicated in Fig.~\ref{fig1}, is attractive to ssDNA. Fig.~\ref{fig1} shows a
typical configuration of the virus surface with adsorbed DNA. Generally,
attractive monomers may form a single attractive patch or be distributed
randomly on the surface, as implemented below. Another possibility is to form a
structure of interconnected ridges mimicking a non-uniform distribution of
attractive capsid proteins on the virus surface. Note that for partially
attractive surfaces, the ssDNA monomers proximal to both attractive and neutral
capsid monomers are equally counted adsorbed by the above criterion.
   
The contractile force $f$ along the DNA axis acting on the dsDNA extremities
connected to the hydrogel is due to both entropic ssDNA fluctuations tending
to coil up the chain and (partial) adsorption to the virus capsid. To find 
$f$ for a fixed length $l_{\text{ss}}$ of ssDNA and for different overall DNA
end-to-end distance $y_0$, we compute the binding energy $U(y_0)$ using the weighted histogram
analysis method (WHAM) \cite{wham}.
The latter is a particular implementation of the umbrella sampling method. Then,
we evaluate the force as
\begin{equation}
f=-\frac{d}{dy_0}[U(y_0)+F_\text{{melt}}(y_0)]=k_y\delta l.
\end{equation}
Here the elastic constant $k_y$ for the displacement $\delta l$ of the DNA ends
varies in the range $k_y\sim0.1\ldots1k_BT/\text{nm}^2$ for different DNA
densities and for the Young modulus of a hydrogel, $E\sim20\text{kPa}$.
This value of
$E$ is due to other polymeric components of the hydrogel supporting the
incorporated, regular lattice of DNA interacting with viral
particles \cite{REM}.

For a typical DNA density ($s\sim70\ldots150\text{nm}$), we compute the
cross-section area
per DNA in our cell model as $S\approx\pi(z_0+R)^2$. Then, the microscopic
contraction of one DNA in the cell via the Young modulus can be related to the
relative contraction of the entire material via
\begin{equation}
\frac{\delta l}{l_{\text{ss}}}\approx\frac{f}{E \pi(z_0+R)^2}.
\label{contraction}
\end{equation}
This contraction is triggered by ssDNA adsorption to the virus surface and
represents the macroscopically measured quantity.

\subsection{Fixed ssDNA length}

We now present the results of our extensive molecular dynamics
computer simulations. To that end we
first obtain the statistical behavior for the DNA-virus interaction for a fixed
length of the available ssDNA fragment, before determining the optimal ssDNA
length self-consistently. We note that the relatively high initiation barrier
$F_s$ for DNA bubble formation prohibits the creation of short molten stretches
of ssDNA. Only longer ssDNA fragments are stabilized by binding to the viral
capsid. 

\begin{figure}
\begin{center}
\includegraphics[width=6.8cm]{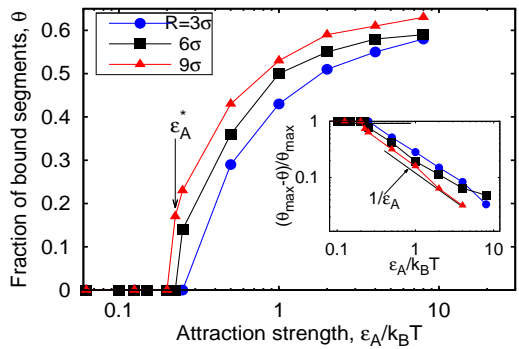}
\end{center}
\caption{Fraction $\theta$ of bound DNA base-pairs as function of ssDNA-virus
attraction $\epsilon_A$ for uniformly attractive viral particle ($p=1$) and
varying capsid radius $R$. Parameters: DNA-virus distance $z_0=8\sigma$, number
of ssDNA monomers $n=101$, ssDNA end-to-end separation $y_0=(n-1)\sigma/2=50
\sigma$. Inset: the residual fraction of non-adsorbed monomers reveals the
scaling $1/\epsilon_A$ (full line). Error bars are comparable to the symbol
size.
\label{fig-theta}}
\end{figure} 

We find that
the fraction $\theta$ of adsorbed segments increases with the adsorption
strength $\epsilon_A$ once a critical value $\epsilon_A^\star$ is exceeded.
After this transition the amount of adsorbed ssDNA segments increases up to a
saturation plateau given by $\theta_{\text{max}}$, see Fig.~\ref{fig-theta}.
As mentioned above, the geometry of our setup influences the shape of this
transition, in particular, the distance $z_0$ between the central
DNA axis and the surface of the capsid, see below. Due to the transition as a
function of the adsorption strength, the standard sigmoidal DNA melting curve
$\theta(T)$ known from thermal melting in our adsorption-induced melting
scenario acquires a kink. According to Fig.~\ref{fig-theta} the transition
of $\theta=\theta(\epsilon_A)$ appears to be of second order. For large
viruses and strong ssDNA-virus attraction, nearly all chain monomers are
adsorbed and non-adsorbed ssDNA fragments are therefore fully stretched.

For a fixed DNA-virus separation $z_0$, the fraction $\theta$ increases with
the capsid radius $R$ as the overall number of available attractive patches
on the viral shell increases. In the limit of strong binding
affinity $\epsilon_A$, the fraction $\theta$ saturates to the geometry-dependent
value $\theta_{\text{max}}(R,z_0)$.
We observe that smaller viral particles naturally require larger
attraction strength for the onset of the denaturation adsorption and yield
smaller fractions of bound ssDNA bases, as shown in Figs.~\ref{fig-theta} and S1
(Supplementary Material).

Interestingly, the decrease of the fraction of non-adsorbed ssDNA bases with
the attraction strength fulfills the scaling relation
\begin{equation}
\frac{\theta_{\text{max}}-\theta}{\theta_{\text{max}}}\sim\frac{k_BT}{
\epsilon_A},
\label{thetamax}
\end{equation}
for sufficiently large values of $\epsilon_A$. This universal dependence is
illustrated in the inset of Fig.~\ref{fig-theta}. The scaling relation
\eqref{thetamax} is fully consistent with the extension of a polymer
chain in a limit of strong stretching forces.
This follows from the relative chain extension $\overline{r}/L_0=\coth[f\sigma
/(k_BT)]-1/[f\sigma/(k_BT)]\approx1-1/[f\sigma/(k_BT)]$ of a worm-like polymer
at large applied forces $f\sigma/(k_BT)\gg1$. Here the chain contour length
is $L_0=(n-1)\sigma$. As the stretching force is due to ssDNA adsorption to
the viral capsid, we have $f=\epsilon_A/\sigma$ and $(L_0-\overline{r})/L_0\sim(
\theta_{\text{max}}-\theta)/\theta_{\text{max}}\sim k_BT/\epsilon_A$.

\begin{figure}
\begin{center}
\includegraphics[width=6.8cm]{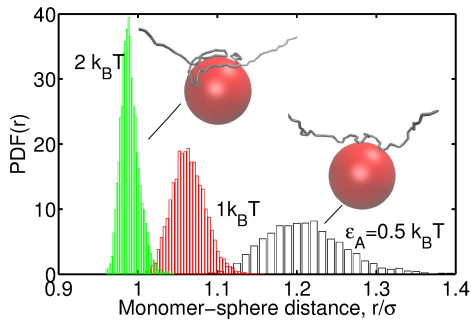}
\end{center}
\caption{Probability density function of the radial distance of ssDNA monomers
from the capsid surface for different values of the adsorption strength
$\epsilon_A$, ranging from weak (black) over intermediate (red) to strong
binding (green). Parameters are the same as in Fig.~\ref{fig-theta}, and $R=6
\sigma$.
\label{fig-PDF}}
\end{figure}

In line with our expectations, the distribution of the chain monomers adjacent
to the attractive viral capsid surface becomes strongly localized at larger
binding affinities, see Fig.~\ref{fig-PDF}. For the case $\epsilon_A=2k_BT$ the
width of this distribution is indeed very narrow. This distribution significantly
broadens and its maximum shifts away from a monomeric distance from the capsid
surface when $|\epsilon_A|$ decreases. This is an obvious tradeoff between
entropic and enthalpic effects when $\epsilon_A$ is comparable to thermal energy.

A natural question to ask is whether, similar to the thermal melting transition,
the sharpness of the binding-induced DNA melting transition increases with DNA
length. We study the effect of ssDNA length in Fig.~S2,
observing that shorter chains require larger adsorption strengths to initiate the
DNA-virus binding (at a constant DNA-virus distance). We also observe faster
saturation to the geometry-limited value $\theta_{\text{max}}$. However, the
transition does not appear sharper for longer ssDNA molecules: approximately
the same number of adsorbed polymer segments is detected at the onset of ssDNA
binding at the critical attraction strength $\epsilon_A=\epsilon_A^\star$. 

\begin{figure}
\begin{center}
\includegraphics[width=6.8cm]{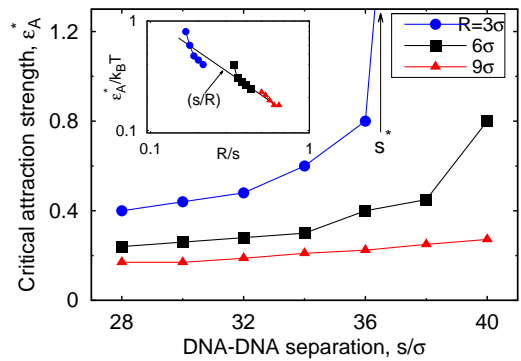}
\end{center}
\caption{Critical adsorption strength $\epsilon_A^{\star}$ for the onset of
ssDNA adsorption, plotted at varying DNA-DNA separation $s$ and virus sizes
$R$. The maximal DNA-DNA separation at which the adsorption strength diverges
is indicated as $s^\star$. Inset: same set of data sampled at constant $s$
values reveal a scaling $\epsilon_A^\star\sim s/R$. Parameters: DNA-shell
distance $z_0=(s-2R)/2$, other parameters as in Fig.~\ref{fig-theta}.
\label{fig-dia}}
\end{figure} 

We also find that the critical value $\epsilon_A^{\star}$ decreases for
larger radii $R$ of the capsid, as shorter DNA-virus separations need to
be bridged by the molten DNAs, thus facilitating the adsorption process, see
Fig.~\ref{fig-dia}. In comparison, the critical $\epsilon_A^{\star}$ for
different $R$ with same $z_0$ is nearly identical (not shown). This indicates
small effects of the shell surface curvature on the onset of the adsorption
transition. For adsorption strengths well above the adsorption transition,
in contrast, the surface curvature effects are vital because they regulate
the number of sites available for adsorption for a given chain length $l_{
\text{ss}}$ and shell-DNA separation $z_0$, as seen from Fig.~\ref{fig-theta}.

\begin{figure}
\begin{center}
\includegraphics[width=6.8cm]{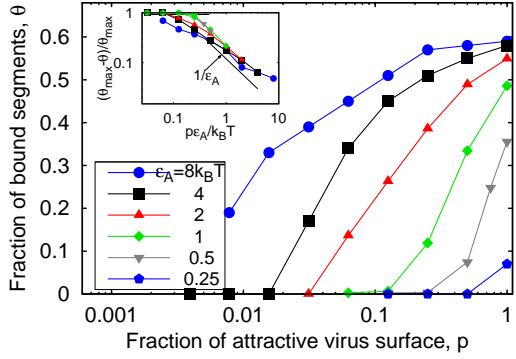}
\end{center}
\caption{Fraction of bound ssDNA segments growing with the virus surface
coverage by attractive monomers, $p$. Inset: fraction of segments remaining
non-bound exhibiting the universal scaling \eqref{thetamax}. Other parameters
as in Fig.~\ref{fig-theta}, and $R=6\sigma$.
\label{fig-p}}
\end{figure}

At a critical DNA-DNA separation $s^{\star}=s^{\star}(R)$ the length of the
ssDNA segment is insufficient to allow adsorption to the capsid. The critical
attraction strength at this adsorption-desorption transition of the ssDNA
naturally
diverges, as shown in Fig.~\ref{fig-dia}. Well above the adsorption transition
(e.g., at relatively small $s$ values) the data reveal a distinct scaling
of the critical adsorption strength with the capsid radius,
\begin{equation}
\epsilon_A^\star(R)\sim s/R=2z_0/R+2.
\end{equation}
This asymptote, shown in the inset of Fig.~\ref{fig-dia}, reveals the geometric
competition between the DNA-virus separation $s$ and the radius $R$.

In Fig.~\ref{fig-p} we present the results for the fraction $\theta$ of bound
ssDNA monomers with varying fraction $p$ of attractive, randomly-distributed
capsid monomers. For larger ssDNA-virus attraction strengths $\epsilon_A$,
smaller $p$ values are sufficient to trigger the DNA adsorption, that is,
the parameter $p$ of the surface coverage by attractive capsid monomers is
complementary to the adsorption energy $\epsilon_A$ per monomer \cite{REMM}.
The DNA adsorption reveals an apparent second-order
continuous transition in the variable $p$. It is thus natural to introduce the
generalized parameter $p\epsilon_A$ and to re-examine relation
\eqref{thetamax} obtained above for the uniformly-attractive sphere. We
observe that the fraction of unbound ssDNA monomers exhibits the analogous
scaling behavior \eqref{thetamax}, substituting $p\epsilon_A$ for
$\epsilon_A$.

\subsection{Optimal ssDNA length and hydrogel contraction}

In the previous subsection, we studied the adsorption transition at a fixed
ssDNA length. Physically, this length is determined self-consistently
from the competition of DNA-virus attraction and DNA melting free energy, as
detailed here.

\begin{figure}
\begin{center}
\includegraphics[width=6.8cm]{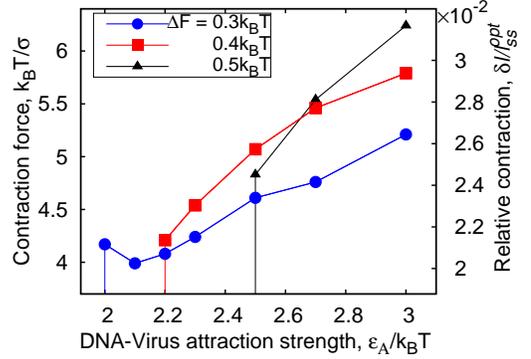}
\end{center}
\caption{Contraction force and the relative hydrogel contraction, per one
DNA bound, as a function of $\epsilon_A$. As $\sigma$=4 nm we have that $k_BT/
\sigma\approx$1 pN. Parameters: $R=6\sigma$, $z_0=8\sigma$, $p=1/3$, and $F_s
=10k_BT$.
\label{fig-contr}}
\end{figure}  
  
As the DNA length $l_{\text{ss}}$ of the overall ssDNA segment is varied, we
find an optimal length $l_{\text{ss}}^{\text{opt}}$, that corresponds to the
minimum of the total free energy. With progressing DNA adsorption the attractive
energy $U$ decreases almost linearly for short ssDNA and appears to saturate for
longer chains. This slower increase of $|U|$ is due to a self-repulsion of the
segments already adsorbed on the sphere. The melting energy cost grows linearly
with the length of molten DNA, see Eq.~\eqref{fmelt} and
Fig.~S3.

In Fig.~S4 we show the optimal ssDNA length for three energy
differences $\Delta F$ between ssDNA and dsDNA. The melting transition only
takes place if the total free energy in the molten-adsorbed state is lower
than that in the helical-desorbed state. This defines the minimal bubble
size when DNA virus-induced melting becomes energetically beneficial. Below a
critical adsorption strength $\epsilon_A^{\star\star}$ , no DNA melting takes
place and $l_{\text{ss}}^{\text{opt}}=0$. For $\epsilon_A>\epsilon_A^{\star
\star}$ the value $l_{\text{ss}}^{\text{opt}}$ starts to grow from a finite
value with $\epsilon_A$, as shown in Fig.~S4. This jump-like behavior
is indicative of a first-order transition, in contrast to the continuous change
in $\theta$ presented in Fig.~\ref{fig-p}. Smaller $\Delta F$ values favor bubble formation and an earlier onset of DNA opening, and thus shift the equilibrium
towards the molten DNA state.

The dependence of the optimal ssDNA length on the model parameters defines
the onset of a longitudinal contraction in the system. Starting from the
adsorption strength $\epsilon_A^{\star\star}$, the contraction force per
DNA increases nearly linearly with the ssDNA-virus adhesion strength,
Fig.~\ref{fig-contr}. The contraction force acting on the DNA ends follows the
linear scaling with $\epsilon_A$ given by Eq.~\eqref{force-contraction}.
For larger $\Delta F$ values the equilibrium length of the ssDNA decreases,
while its effect on the gel contraction becomes stronger.

Collecting our previous results we arrive at the main conclusion of this work,
the relative contraction $\delta l/l_{\text{ss}}$ of the hydrogel. According to
Eq.~\eqref{contraction}, this relative contraction $\delta l/l_{\text{ss}}$
scales linearly with the contraction force $f$ and is inverse-proportional to the
Young modulus $E$. According to Fig.~\ref{fig-contr}, the force $f$ is 
approximately proportional to the attraction strength $\epsilon_A$, thus
confirming Eq.~\eqref{force-contraction}. The right axis in Fig.~\ref{fig-contr}
shows the corresponding relative contraction $\delta l/l_{\text{ss}}$ effected
by a single ssDNA-virus interaction within one of the `elementary cells' of our
setup. Accordingly, for the value $E=20\text{kPa}$ of the hydrogel Young modulus,
this would give rise to a contraction of $3\ldots5\%$. Assuming that at least
two ssDNA per `elementary cell' contribute, the contraction of the entire
hydrogel would reach almost 10\%. This, however, is a conservative estimate
with respect to the relatively high value for the Young modulus $E$ that we
chose here. Using softer hydrogels, $E$ can easily be reduced considerably,
thus leading to a much stronger response to the viral binding. This macroscopic
hydrogel contraction is thus immediately detectable.

\section{Conclusion}

Based on extensive computer simulations and statistical analyses we explored the
thermodynamic properties of DNA melting induced by binding of single-stranded
DNA to viruses and the effected longitudinal tension build-up in the involved
DNA chains. In particular, we determined how a preferable adhesion of ssDNA
fragments to the surface of the virus shifts the DNA melting equilibrium and
alters the character of the DNA melting transition. We found the critical
adsorption strength of ssDNA chains to the attractive viral shell under the
conditions of DNA confinement in an elastically-responsive polymer matrix of a
hydrogel. The statistical properties of the partially molten and partially
adsorbed DNA chains obtained from computer simulations are in line with the
theoretical expectations. We found several scaling relations for the fraction of
adsorbed DNA, the critical adsorption strength, and, most importantly, the
relative shrinkage of the suspending hydrogel.

As a proof of concept we demonstrated that the presence of viruses in a liquid
solution may be detected directly by mechanical response of a hydrogel,
transduced by pre-stretched DNA chains suspended in the hydrogel. These DNAs
partially melt and thus expose single-stranded DNA, that itself binds to the
virus shell. This binding can, in principle, be made specific for certain types
and sizes of viruses. All ingredients are fairly inexpensive to produce, and such
a setup can easily be miniaturized. We believe that this setup is a good candidate
for mobile use or for use at local health care centers. Alternative to measuring
the mechanical deformation of the hydrogel, one could also envision optical
signals due to either DNA melting (similar to molecular beacon setups
\cite{beacon}) or strong local hydrogel deformation. The binding to DNA located
in separate domains of the hydrogel can be made specific to certain virus types
by biochemically specific functionalized units on the virus.

We believe that the effects determined herein may form an alternative basis for
miniaturized and inexpensive viral detection under non-clinical conditions.

\begin{acknowledgments}
The authors acknowledge funding from the Academy of Finland (FiDiPro scheme
to RM), the German Research Council (DFG Grant CH 707/5-1 to AGC), and the
German Ministry for Education and Research.
\end{acknowledgments}

\begin{appendix}

\section{Simulations scheme}
\label{appa}

We consider the adsorption of ssDNA onto an attractive surface of a viral
particle in a simplified model. The ssDNA is modeled by Lennard-Jones
(LJ) particles inter-connected by finitely extensible nonlinear elastic
(FENE) springs of the potential
\begin{equation}
U_{\text{FENE}}(r)=-\frac{k}{2}r_{\text{max}}^2\ln\left(1-\frac{r^2}{
r_{\text{max}}^2}
\right),
\end{equation}
where $k$ is the FENE spring constant and $r_{\text{max}}$ is the maximum
allowed separation between neighboring monomers. Excluded-volume
interactions between the monomers are given by the short-ranged truncated LJ
repulsion,
\begin{equation}
U_{\text{LJ}}(\epsilon,r)=\left\{\begin{array}{ll}
4\epsilon[(\sigma/r)^{12}-(\sigma/r)^6]+\epsilon,&r<2^{1/6}\sigma\\[0.2cm]
0,&\text{otherwise}\end{array}\right.
\end{equation}
Here, $r$ is the monomer-monomer distance, $\sigma$ is the monomer diameter,
and $\epsilon$ is the strength of the potential. We set the parameters to
$k=30$, $r_{\text{max}}=1.5\sigma$, and $\epsilon=1$. The monomer size
is $\sigma=4 $nm, close to the Kuhn length of a highly flexible ssDNA. In
contrast to dense DNA fibers \cite{nords94}, here we can neglect DNA-DNA
interactions because the intermolecular separations exceed the diameter of
a typical virus, $2R\sim$50-100 nm, and the electrostatic forces between
individual DNA chains are well screened.

Performing simulations of a free chain, we checked that the effective ssDNA
persistence length in this model agrees well with the known estimate of
$l_p=$1-4 nm \cite{chen,rechendorff}, and the chain gyration radius follows
the Flory $R_g^2\sim n^{3/5}$ law for an excluded-volume/self-avoiding polymer
as function of the number $n$ of monomers.

The spherical viral shell of radius $R$ is modeled by a densely packed assembly
of LJ beads of size $\sigma$. A finite fraction $0<p<1$ of the monomers,
distributed either in a random or ordered fashion on the viral surface, is
attractive for ssDNA segments in their vicinity. The ssDNA-virus attraction is
modeled via the same spherically symmetric LJ potential,
\begin{equation}
U_{\text{PV}}(r)=U_{\text{LJ}}(\epsilon_A,r),
\end{equation}
but with a larger cutoff-distance $r^{\star}=2.5\sigma$ to ensure the existence
of an attractive potential branch at larger distances and with varying attraction
strength $\epsilon_A$.
The dynamics of the $i$th chain monomer is described in our molecular dynamics
analysis by the Langevin equation
\begin{eqnarray}
\nonumber
m\frac{d^2\mathbf{r}_i(t)}{dt^2}&=&-\xi\mathbf{v}_i(t)+\mathbf{F}_i^R(t)-\sum_{
j,j\neq i}\nabla\Big(U_{\text{LJ}}(\mathbf{r}_i-\mathbf{r}_j)\\
&&\hspace*{-0.8cm}
+U_{\text{FENE}}(\mathbf{r}_i-\mathbf{r}_j)+U_{PV}(\mathbf{r}_i-\mathbf{r}_V)
\Big),
\end{eqnarray}
where $m$ is the monomer mass, $\xi$ the friction coefficient, $\mathbf{v}_i$
the monomer velocity, and $\mathbf{F}_i^R$ represents zero-mean Gaussian noise
with component-wise $\delta$-correlation, $\left<\mathbf{F}_i^R(t)\mathbf{F}_j^R
(t')\right>=6\xi k_BT\delta_{i,j}\delta(t-t')$. The positions of the two polymer
ends, $(0,0,z_0)$ and $(0,y_0,z_0)$, and of the virus center $\mathbf{r}_V=(0,
y_0/2,-R)$ are fixed, see Fig.~1. In the simulations, we set
$m=1$, $\xi=1$, and $k_BT=1$. The equation of motion is integrated using the
velocity Verlet algorithm \cite{verlet} with a time step $\Delta t=0.01$.

\end{appendix}


\begin{thebibliography}{99}
\bibitem{REM} For many hydrogels the stress-strain relations and the
Young modulus are obtained from analysis of elastic deformations and ball
indentation \cite{hdelast1}. The Young modulus varies in the range $E=0.01
\ldots30\text{kPa}$, depending on the gel cross-links density, its chemical
composition, and temperature \cite{okay13}. Typically, the modulus grows
nearly linearly with concentration of cross-linked polymer chains. Typical
values of 5-20 kPa are reported for the aragose gels at 0.5-1.5\% polymer
concentrations in the regime when gel stretching dominates over its
bending. The compositions of hydrogels can include various polyelectrolyte
chains in different proportions, including a double-stranded DNA (dsDNA).
\bibitem{REMM} Some deviations from this scaling for strong
adsorption are due to the fact that in this case only a small number of polymer
monomers are actually in tight contract with the surface. The monomers in
polymers loops interconnecting these adsorbed links are however also counted as
adsorbed as they are close to the surface. This gives rise to a higher fraction
of adsorbed monomers and some deviation from the universal behavior
\eqref{thetamax} in the
strong adsorption limit. Smaller $p$ values and larger adsorption strength
are more efficient in binding ssDNAs to the capsid shells.
\end{thebibliography}
\end{document}